\documentclass[10pt]{article}
\usepackage{spconf,amsmath,graphicx}
\usepackage{graphicx}
\usepackage{booktabs}
\usepackage{environ}
\usepackage{amsmath}
\usepackage{enumitem}
\usepackage{scrextend}

\NewEnviron{FitToWidth}[1][\textwidth]{%
\begin{center}
  \makebox[#1]{\resizebox{#1}{!}{\BODY}}%
\end{center}
}

\let\OLDthebibliography\thebibliography
\renewcommand\thebibliography[1]{
  \OLDthebibliography{#1}
  \setlength{\parskip}{0pt}
  \setlength{\itemsep}{0pt plus 0.3ex}
}
\title{Analysis of Self-Supervised Speech Models on Children's Speech and Infant Vocalizations}
\name{Jialu Li$^{1,2}$, Mark Hasegawa-Johnson$^{1,2}$, Nancy L. McElwain$^{2,3}$}
\address{
  $^1$Department of Electrical and Computer Engineering, University of Illinois\\ 
  $^2$Beckman Institute for Advanced Science and Technology, University of Illinois \\
$^3$Department of Human Development and Family Studies, University of Illinois\\
\texttt{\{jialuli3, jhasegaw, mcelwn\}@illinois.edu}
}

\begin{document}
\ninept
\maketitle
\begin{abstract}
To understand why self-supervised learning (SSL) models have empirically achieved strong performances on several speech-processing downstream tasks, numerous studies have focused on analyzing the encoded information of the SSL layer representations in adult speech. Limited work has investigated how pre-training and fine-tuning affect SSL models encoding children's speech and vocalizations. In this study, we aim to bridge this gap by probing SSL models on two relevant downstream tasks: (1) phoneme recognition (PR) on the speech of adults, older children (8-10 years old), and younger children (1-4 years old), and (2) vocalization classification (VC) distinguishing cry, fuss, and babble for infants under 14 months old. 
For younger children's PR, the superiority of fine-tuned SSL models is largely due to their ability to learn features that represent older children's speech and then adapt those features to the speech of younger children. For infant VC, SSL models pre-trained on large-scale home recordings learn to leverage phonetic representations at middle layers, and thereby enhance the performance of this task.
\end{abstract}
%

%
\section{Introduction}
\label{sec:intro}

Automatically analyzing children's speech and vocalizations, such as automatic speech recognition (ASR) and vocalization classification (VC),  could yield significant advantages in the fields of education, healthcare, and developmental psychology. For example, automatic analysis of children's speech can assist both clinicians and parents in identifying autism~\cite{gong2018automatic}, speech and language disorders~\cite{toki2021game}, allowing for timely intervention. Additionally, children's ASR plays an important role in developing interactive educational software that can adapt to the child's speech and language abilities~\cite{hagen2003children}, which provides opportunities to personalize feedback and increase the efficiency of educational programming and support.

One major obstacle to developing children's speech analysis models is the sparsity of available labeled datasets. Recently, self-supervised learning (SSL) models, such as wav2vec 2.0 (W2V2)~\cite{wav2vec} and HuBERT~\cite{hsu2021hubert}, have significantly advanced children's speech technology with limited labeled data. SSL model first uses pre-training on a large amount of unlabeled audio data. By fine-tuning a pre-trained SSL on a small amount of labeled data, it excels on multiple downstream tasks. With SSL, several studies have reported strong performances on children's ASR~\cite{fan22d_interspeech,9864219,10122501}, child-adult speaker recognition~\cite{lahiri23_interspeech}, infant cry detection~\cite{Gorin2023SelfSupervisedLF}, and children's VC task~\cite{xu23e_interspeech,li23e_interspeech}. 

To understand why SSL models empirically have achieved strong performances on downstream tasks, previous studies have examined the encoded information in SSL layer representations. Pasad et al.~\cite{pasad2021layer} performed layer-wise analysis on SSL models and found acoustic and linguistic properties were encoded in several layers, and pre-trained W2V2 model behaved similarly to an ``autoencoder'' style. 
Li et al.~\cite{li2023exploration} similarly explored W2V2 fine-tuned on emotional corpora using CCA and discovered W2V2 discards paralinguistic information that is irrelevant for word recognition task. 
Other studies expanded layer-wise analysis for accent~\cite{yang23v_interspeech} and prosody~\cite{lin2023utility} probing. 

Previous studies have mainly focused on analyzing and understanding the performances of SSL models involving adult speech. Little work has been conducted in probing SSL models for encoding children's speech and vocalizations, particularly for younger children under the age of 5. In one relevant study, Lavechin et al.~\cite{lavechin23_interspeech} created a benchmark of children's language acquisition to probe SSL-based spoken language model on lexical and syntactic levels. Yeung et al.~\cite{yeung2018difficulties} revealed that the ASR trained on kindergarten speech performed much worse than the ASR trained on 1st-grade speech when tested on kindergarten speech due to the large variability of kindergarten phonetics, and 1st-grade children's speech was perhaps useful for training a kindergarten ASR. Other studies~\cite{shetty23_interspeech,potamianos2007review} have analyzed and showed variability of children's acoustic features across ages.
In this study, we analyze how SSL models encode children's speech and vocalizations on two tasks: (1) phoneme recognition (PR) for the speech of adults, older children (8-10 years old), and younger children (1-4 years old), and (2) VC distinguishing \textit{cry}, \textit{fuss}, and \textit{babble} for infants under 14 months.  Code and model weights are available\footnote{https://huggingface.co/lijialudew/}.
Our detailed findings are listed as the following:
\begin{itemize}[noitemsep,leftmargin=*]
\item For PR, SSL models fine-tuned on younger children's speech maintain older children's phonetics but lose most of adult phonetics. SSL models pre-trained on children's speech under 5 years old shows limited encoding of the phonetics specific to this age group.
\item For infant VC, SSL model pre-trained on home recordings achieves superior performance by utilizing representations of middle layers, which are encoded with higher-level phonetic features. SSL model pre-trained on adult speech uses lower layers that are close to input acoustic features, which results in inferior performance. SSL representations are mostly correlated with paralinguistic feature groups of energy, mel-frequency cepstral coefficients (MFCC), and pitch. 
\end{itemize}


\section{Phoneme Recognition Analysis}
\label{sec:format}
\subsection{Data}
We use three datasets for PR analysis, including LibriSpeech \cite{libri}, My Science Tutor (MyST)~\cite{ward2011my}, and Providence~\cite{demuth2006word}. 
For LibriSpeech, we use train-clean-100 (100-hour speech with 251 speakers), dev-clean (5.4-hour speech with 40 speakers), and test-clean (5.4-hour speech with 40 speakers) for training and testing adult speech.
MyST contains conversational speech of 1371 students between third and fifth grades ($\sim$8-10 years old) with a virtual tutor. We select short transcribed utterances (\textless 15s) for training and testing to align with shorter vocalizations produced by younger children. We have 84.2h/14.1h/15.1h for training/development/testing sets respectively. Providence contains longitudinal audio and video recordings of six English-speaking children (3 boys: Alex, Ethan, and William, and 3 girls: Lily, Naima, and Violet) from ages 1 to 4 interacting with their mothers at home. Annotators transcribed children's speech using SAMPA phonetic symbols. We manually filter out some of the highly noisy recordings and use transcribed child audio for fine-tuning SSL models. In total, we train on 84.0h utterances of four children (Ethan, William, Lily, and Naima) and test on 24.8h utterances of the other two (Alex and Violet). To perform PR, we convert transcripts of all three datasets into international phonetic alphabet (IPA) format. For LibriSpeech and MyST, we use eng\_to\_ipa software\footnote{https://pypi.org/project/eng-to-ipa/}. For Providence, we directly map SAMPA symbols to IPA. In total, we have 53 unique IPA phones.
\subsection{Experimental Setting}
We build PR models using W2V2 and HuBERT on SpeechBrain~\cite{ravanelli2021speechbrain} framework. The training objective for all fine-tuned SSL models is to minimize phone-level CTC loss~\cite{graves2006connectionist}. For inference, we use CTC greedy decoding without any language model. For brevity, we use the format \textit{W2V2/HuBERT-Dataset} to denote SSL models that are either pre-trained or fine-tuned on specific \textit{Dataset} using \textit{W2V2} or \textit{HuBERT}.  

\textbf{\textit{W2V2/HuBERT-Libri100h}}: We add an output linear layer of hidden node 384 followed by CTC loss on top of \textit{\textbf{W2V2/HuBERT-Base}}, W2V2/HuBERT pre-trained on 960-hour unlabeled LibriSpeech, consisting of 12 transformer layers with a hidden dimension of 768. We first freeze \textit{W2V2/HuBERT-Base} and only fine-tune the added linear layer for 90 epochs using LibriSpeech with phone-level CTC loss; then we fine-tune the entire model for an additional 10 epochs until convergence. \textbf{\textit{W2V2/HuBERT-MyST}}: We keep the linear layer trained on \textit{W2V2/HuBERT-Libri100h} and continue fine-tuning the entire model using MyST for another 10 epochs until convergence. To build PR model for Providence, we test both two-level (\textbf{\textit{W2V2/HuBERT-Libri100-Pro}}) and three-level fine-tuning (\textbf{\textit{W2V2/HuBERT-MyST-Pro}}). For both cases, we keep the linear layer trained on \textit{W2V2/HuBERT-Libri100} or \textit{W2V2/HuBERT-MyST} and then fine-tune the entire model for another 10 epochs using Providence. 
We also explore adding a linear layer on (1) \textit{W2V2/HuBERT-Base}, and (2) \textbf{\textit{W2V2-LL4300h}}~\cite{li23e_interspeech}, a W2V2 model pre-trained on 4300-hour unlabeled home recordings used in our past study. We collected these recordings from 245 families of children under 5 years old via infant-wearable recording devices, LittleBeats (LB)~\cite{mcelwain2023evlauating} and LENA\footnote{LB and LENA data collection and annotation was supported by the funding from the National Institute on Drug Abuse (R34DA050256), National Institute of Mental Health (R21MH112578), and USDA National Institute of Food and Agriculture  (ILLU-793-368).}~\cite{gilkerson2008lena}.
We directly fine-tune \textit{W2V2/HuBert-Base} or \textit{W2V2-LL4300h} with the linear layer using Providence. For all experiments, learning rates (LR) are set for Adam optimizer as 9e-3 and 1e-4 for the output linear layer and W2V2/HuBERT respectively; scheduler with the new-bob technique adjusts LR based on the development set performance after each epoch. Epoch with the best performance of the development set is used for final evaluation.
\subsection{Canonical Correlation Analysis for Phoneme Labels}
\label{sec:CCA_scores}
Canonical correlation analysis (CCA)~\cite{pmlr-v28-andrew13} is a statistical method for finding the maximum correlation between linear projections of two sets of continuous-valued vectors. In this study, we use projection-weighted CCA~\cite{Morcos2018InsightsOR}, as described in~\cite{pasad2021layer}, to examine the similarity between SSL layer representation and one-hot phonemes labels for all three datasets.
For extracting phoneme representations from SSL models, we use Montreal Forced Aligner~\cite{mcauliffe2017montreal} to compute phoneme timestamps. For each phoneme, we average its SSL representation within the central third timespan. We discard the first and last third timespans to alleviate co-articulation effects. In total, we extract roughly 30k phonemes, with about 600 samples per phoneme, for each dataset. The final CCA scores are calculated with a cross-validation scheme. Total samples are divided into ten folds. One fold is used for testing, while the remaining nine are for training. The average CCA score from three different test folds is reported.

\begin{table}[]
\setlength{\tabcolsep}{2.0pt}
\resizebox{\columnwidth}{!}{%
\begin{tabular}{l|cc|cc|cc}
\toprule
Dataset & \multicolumn{2}{l}{LibriSpeech} & \multicolumn{2}{l}{MyST} & \multicolumn{2}{l}{Providence}        \\
Model                                         & W2V2          & HuBERT          & W2V2       & HuBERT      & W2V2              & HuBERT            \\
\midrule
\textit{Libri100}                                      & 2.9           & 3.2             & 19.7       & 20.5        & \textgreater{}100 & \textgreater{}100 \\
\textit{MyST}                   & 6.4           & 5.8             & 10.7       & 10.5        & \textgreater{}100 & \textgreater{}100 \\
\textit{Libri100-Pro}                    & 99.3          & 99.1            & 81.8       & 75.6        & 59.9              & 60.2             \\
\textit{MyST-Pro}& 99.1          & 99.4            & 70.8       & 64.1        & 59.6 (*)             & 59.1 (***)              \\
\midrule
\textit{LL4300-Pro}                      & --            & --              & --         & --          & diverged          & -          \\
\textit{Base-Pro}                        & --            & --              & --         & --          & diverged          & diverged        \\
\bottomrule
\end{tabular}%
}
\caption{Phone error rate, in percent, for SSL models evaluated on the test set of LibriSpeech, MyST, and Providence. Results of MAPSSWE tests for \textit{W2V2/HuBERT-Libri100-Pro} and \textit{W2V2/HuBERT-MyST-Pro} models are indicated with * in parentheses. }
\label{tab:phoneme_recognition}
\end{table}

\subsection{Results \& Analysis}

\begin{figure}[t]
  \centering
  \includegraphics[width=0.9\linewidth]{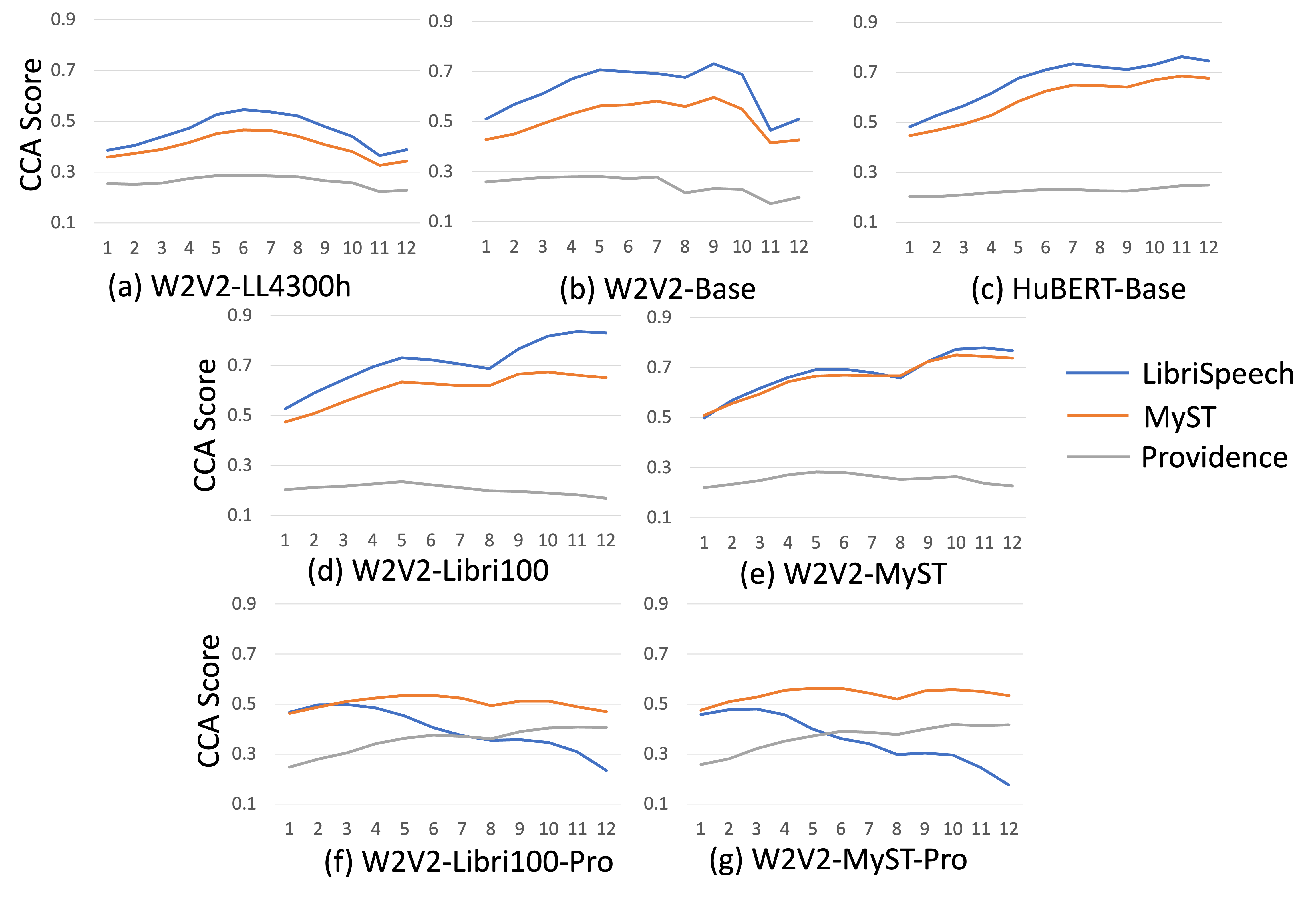}
\caption{CCA scores (y-axis) over 12 layers (x-axis) computed on LibriSpeech, MyST, and Providence datasets for three pre-trained models, (a)-(c), and four fine-tuned W2V2 models, (d)-(g). }
  \label{fig:cca_phoneme_scores}
\end{figure}

Table~\ref{tab:phoneme_recognition} presents phone error rates (PER) for each fine-tuned model evaluated on the testing set for LibriSpeech, MyST, and Providence datasets. Figure~\ref{fig:cca_phoneme_scores} shows the layer-wise CCA scores that can be used to explain performances in Table~\ref{tab:phoneme_recognition}. 

\noindent\textbf{Analysis of Pre-trained Models} For two pre-trained W2V2 models (\textit{W2V2-Base} and \textit{W2V2-LL4300h}) with the pre-training objective as reconstructing masked acoustic input, we see an ``auto-encoder style'' behavior with middle layers (6-10) encoding the richest phonetic information, while top layers exhibit a decline in phonetic content on the speech of adults and older children (Fig~\ref{tab:phoneme_recognition} (a) and (b)). CCA scores on \textit{W2V2-LL4300h} are lower than \textit{W2V2-Base}, which suggests the far-field adults' speech in noisy home recordings may cause degradation in learning phonetics. For \textit{Hubert-Base} (Fig~\ref{tab:phoneme_recognition} (c)), top layers embed the highest phonetic content maybe because its pre-training objective is to predict discrete phonetic units learned from clustering. 
However, all three pre-trained models fail to encode phonetics for younger children, which explains why fine-tuning younger children's speech on these pre-trained models diverged (\textit{W2V2-LL4300h-Pro} and \textit{W2V2/HuBERT-Base-Pro} in Table~\ref{tab:phoneme_recognition}).    

\noindent\textbf{Analysis of Fine-tuned Models} For fine-tuned SSL models, we observe similar trends of CCA scores of HuBERT and W2V2, so we only present the W2V2 set for brevity\footnote{\label{ft1} We present complete sets of CCA scores here: https://sites.google.com/view/ssl-analysis-children-speech. This link also includes supplementary results of sample T-SNE~\cite{van2008visualizing} plots to demonstrate how phoneme clusters of different age groups evolve across layers for fine-tuned SSL models.}. 
Top layers of \textit{W2V2-Libri100h} and \textit{W2V2-MyST} fine-tuned with phonetic targets encode the most phonetic information (Fig~\ref{tab:phoneme_recognition} (d) and (e)), and \textit{W2V2-MyST} reduces the domain shift between the phonetics of adults and older children for all layers to achieve optimal result on MyST corpus. 
We compare PER for \textit{W2V2/HuBERT-Libri100h-Pro} and \textit{W2V2/HuBERT-MyST-Pro} on Providence using statistical significant tests~\cite{nist}, MAtched Pairs Sentence-Segment Word Error (MAPSSWE). We obtain statistically significant better PER using three-level fine-tuning \textit{W2V2-MyST-Pro}, with \textit{p-value=0.018}, and \textit{HuBERT-MyST-Pro}, with \textit{p-value$<$0.001}, than their corresponding two-level fine-tuning models. 
For \textit{W2V2-Libri100h-Pro} and \textit{W2V2-MyST-Pro} (Figure~\ref{tab:phoneme_recognition} (f) and (g)), CCA scores show a smaller gap between phonetics of older and younger children across all layers, but a steady decrease with adult phonetics beyond layer 3. 
To recognize younger children's phonetics through fine-tuning, the SSL model and the output linear layer need to first encode the phonetics of adults or older children as a foundational basis.
Note \textit{W2V2-MyST-Pro} has higher CCA scores of older children's speech than \textit{W2V2-Libri100h-Pro} for top layers, which corresponds to lower PER on MyST on three-level fine-tuning models (\textit{W2V2/HuBERT-MyST-Pro}) than those of two-level fine-tuning (\textit{W2V2/HuBERT-Libri100h-Pro}).


\section{Infant Vocalization Analysis}
\subsection{Data \& Methodology}
We follow our previous work~\cite{li23e_interspeech} on data preprocessing, model architecture, and experimental settings for analyzing infant vocalizations (under 14 months) collected via LB. We separate LB home recordings into 10-min segments for annotation.
Infant vocalizations were manually labeled as \textit{cry}, \textit{fuss}, and \textit{babble}.
In total, we have 70 labeled 10-mins recordings from 22 families. 
We divide the LB audio stream into 2-second intervals of every 0.2 seconds, and the vocalization type is determined as the majority type of centered 1 second.
We follow the leave-one-family-out data partition scheme and have 52/6/12 10-mins recordings from 15/3/4 families for training/development/testing sets, respectively.
In our past study, we used W2V2 models to perform three tasks, including speaker diarization, infant VC, and adult VC tasks. For each task, we employed a weighted average (WA) layer to learn the weights of mean-pooled utterance-level representations from all 12 layers. Then we input the weighted sum features into a feed-forward network (FFN) to obtain classification results. The training objective is the average of cross-entropy losses over three tasks. In this study, we focus on analyzing two models, \textit{W2V2-Base} and \textit{W2V2-LL4300h}, on the infant VC task. We either freeze W2V2 and only fine-tune the WA and FFN or fine-tune the entire model. 
For each experiment, we run each model for 10 epochs and use the epoch with the best performance on the development set for testing.
We further examine the weights learned from the WA layer and rerun the experiments with the best three layers.
\begin{figure}[]
  \centering
  \includegraphics[width=1.0\linewidth]{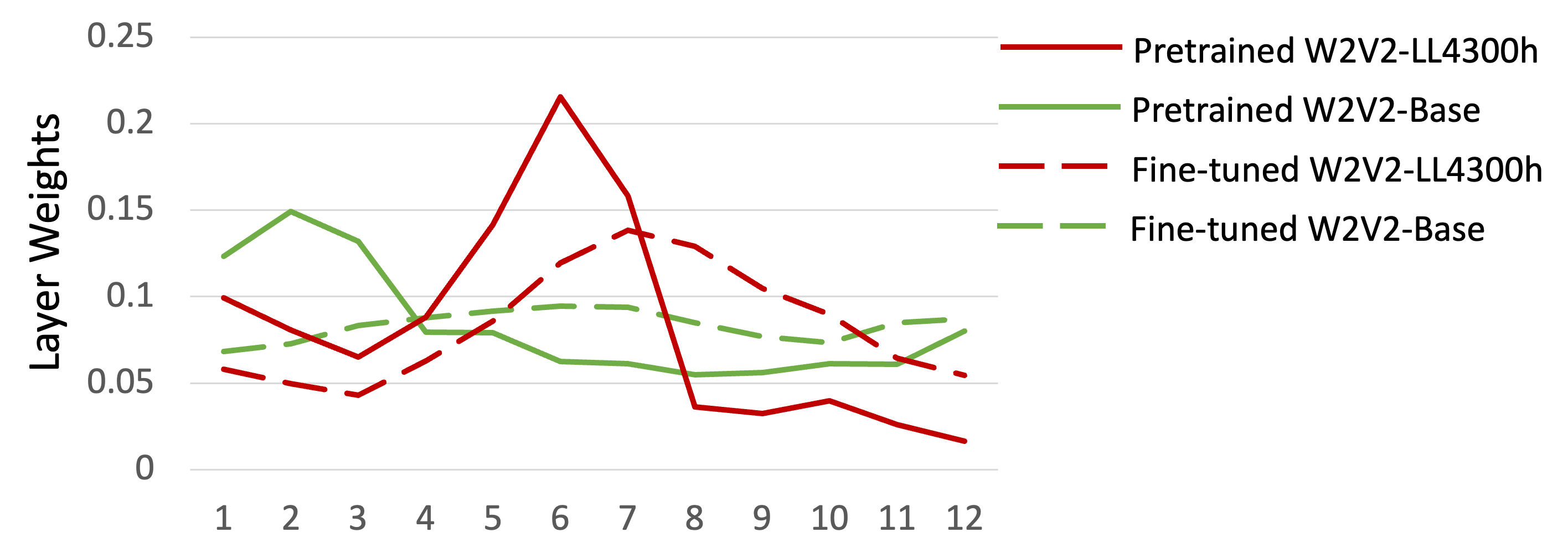}
\caption{Weights learned (y-axis) for 12 layers (x-axis) in weighted average layer for four W2V2 models. For each model, weights sum up to 1 across all layers.}
  \label{fig:CHN_weighted_avg}
\end{figure}
\begin{table}[]
\centering
\scalebox{0.8}{
\begin{tabular}{l|l|l|ll}
\toprule
Model        & PT/FT & Layers & \multicolumn{2}{l}{Infant VC} \\
             &       & Used       & Acc           & F1            \\
\midrule
W2V2-Base    & PT    & All 12 & 68.7          & 49.5          \\
             & FT    & All 12 & 65.7          & 53.5          \\
             \midrule
W2V2-LL4300h & PT    & All 12 & 76.0          & 64.3          \\
             & PT    & 5-7    & 75.9          & 66.4          \\
             & FT    & All 12 & 78.2          & 68.9          \\
             & FT    & 5-7    & 75.2          & 65.1         \\
             \bottomrule
\end{tabular}
}
\caption{Accuracy and unweighted F1 scores for infant VC task. PT=Freeze pre-trained W2V2 and fine-tune output FFN only, and FT=Fine-tune entire model.}
\label{tab:CHN_results}
\end{table}
\subsection{Infant VC Results \& Analysis}
Results of infant VC are shown in Table~\ref{tab:CHN_results}. Figure~\ref{fig:CHN_weighted_avg} shows the weights learned from the WA layer across models. \textit{W2V2-LL4300h} outperformed \textit{W2V2-Base} by a large margin. For \textit{W2V2-Base} pre-trained on adult speech, only lower layers 1-3 are mostly helpful, which are close to input acoustic features. This explains why \textit{W2V2-Base} produces degraded results. By fine-tuning \textit{W2V2-Base} with infant vocalizations, the learned weights become relatively uniform across layers, possibly due to the domain shift involved in learning infant acoustics for the entire model.
For \textit{W2V2-LL4300h} pre-trained on home recordings, we observe layers 5-7 have the largest weights. Interestingly, these layers also heavily encode phonetics of adults and older children (Fig~\ref{fig:cca_phoneme_scores} (a)). By freezing \textit{W2V2-LL4300h} and fine-tuning output FFN only with a weighted sum of layers 5-7, we obtain a 2.1\% gain in terms of the F1 score compared to using a weighted sum of all 12 layers.  
Fine-tuning the entire \textit{W2V2-LL4300h} with a weighted sum of all 12 layers leads to the optimal performance, where layers 6-8 contribute the most weights.

\begin{figure}[t]
  \centering
  \includegraphics[width=0.9\linewidth]{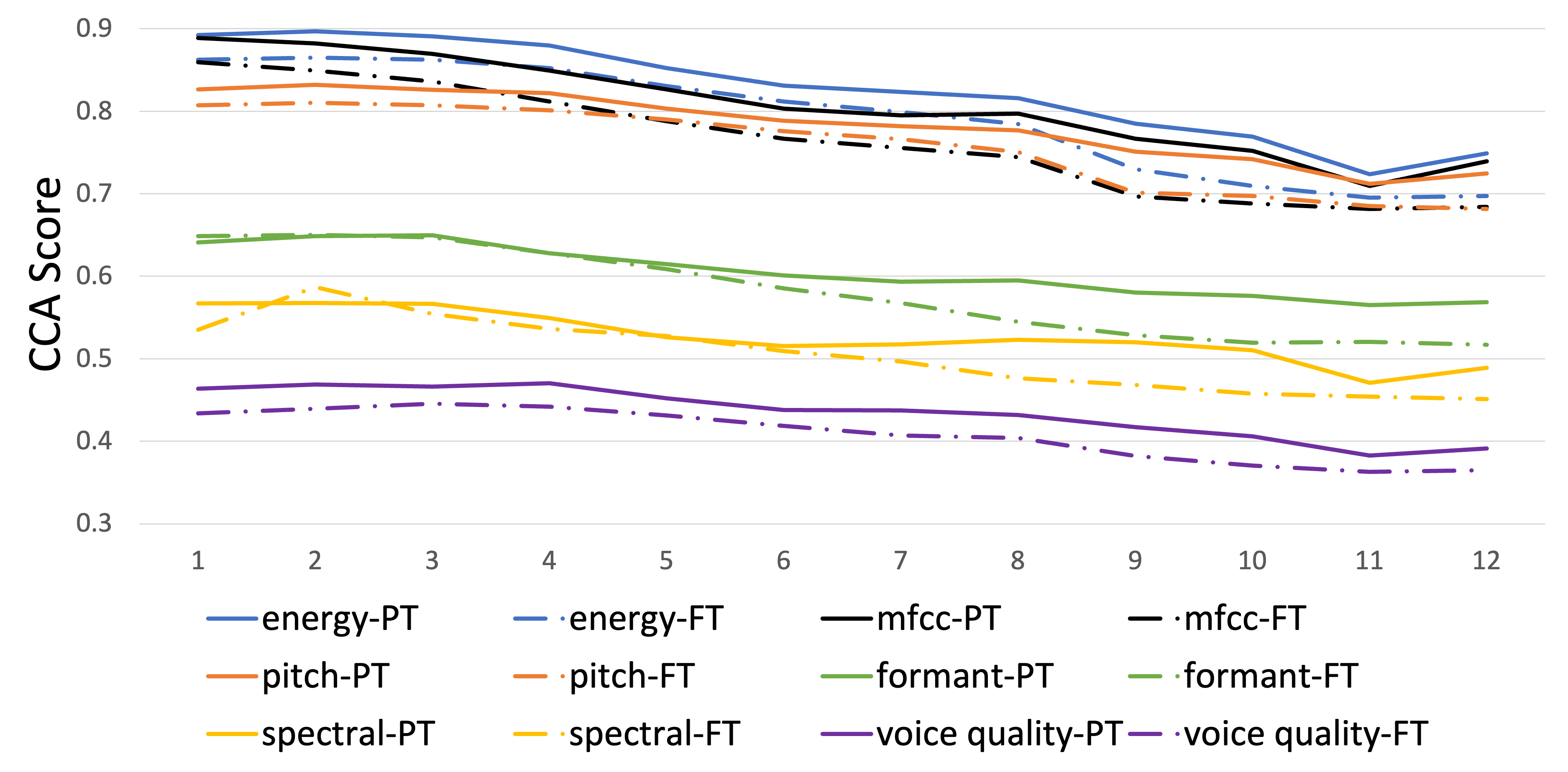}
\caption{CCA scores (y-axis) over 12 layers (x-axis) for paralinguistic feature groups computed on \textit{W2V2-LL4300h}. PT=Pre-train and FT=Fine-tune.}
  \label{fig:CCA_paralinguistic}
\end{figure}
\begin{table}[]
\smallskip
\centering
\setlength{\tabcolsep}{2.0pt}
\resizebox{0.8\columnwidth}{!}{%
\begin{tabular}{l|l|l}
\toprule
Feature Group     & Dim & Low-level Descriptors  \\
\midrule
energy        & 2         & Loudness; Harmonics-to-Noise Ratio                                                                                                                                                                                   \\
\midrule
MFCC          & 4         & Mel-Frequency Cepstral Coefficients 1-4                                                                                                                                                                              \\
\midrule
pitch         & 1         & Pitch                                                                                                                                                                                                                \\
\midrule
formant       & 6         & Formant 1, 2, and 3 frequency and bandwidth                                                                          \\
\midrule
spectral      & 10        & \begin{tabular}[c]{@{}l@{}}Alpha Ratio 50-1000Hz and 1-5k Hz; \\ Hammarberg Index;\\ Spectral Slope 0–500 Hz and 500–1500 Hz;\\ Formant 1, 2, and 3 relative energy;\\ Harmonic difference H1–H2 and H1-A3\end{tabular} \\
\midrule
voice quality & 2         & Shimmer; Jitter    \\
\bottomrule
\end{tabular}%
}
\caption{Paralinguistic feature groups and their dimensions.}

  \label{tab:para_table}
\end{table}

\subsection{CCA for Paralinguistic Features}
We measure the similarity between \textit{W2V2-LL4300h} layer representations and paralinguistic features using CCA. For extracting paralinguistic vectors, we randomly select 4500 2s samples, with 1500 samples each from \textit{cry}, \textit{fuss}, and \textit{babble}. We compute CCA scores cross-validation, similarly to the procedures as described in section~\ref{sec:CCA_scores}.
We use OpenSMILE~\cite{eyben2010opensmile} toolkit to extract Geneva Minimalistic Acoustic Parameter Set~\cite{eyben2015geneva} with 25 low-level descriptors, and we group them into six categories as shown in Table~\ref{tab:para_table}. 
Figure~\ref{fig:CCA_paralinguistic} shows the CCA scores.
W2V2 layer representations are mostly correlated with energy, MFCC, and pitch and less correlated with formant, spectral, and voice quality parameters. 
Pitch is known as a key feature in distinguishing infant vocalizations~\cite{fox1990analysis}, which is also preserved in W2V2 representations.
CCA scores for paralinguistic features generally showed a steady decrease trend from bottom layer 1 to top layer 12. We suspect that this may be due to paralinguistic features being less useful in reconstructing masked input regions in the W2V2 pre-training stage at top layers.
CCA scores for fine-tuned W2V2 are generally lower than pre-trained W2V2 beyond layer 6, which suggests fine-tuned W2V2 has less emphasis on paralinguistic features.



\section{Conclusion}
\label{sec:conclusion}
In this study, we probe several pre-trained and fine-tuned SSL models on two tasks, children's PR and infant VC. For PR, we provide insights into how fine-tuning affects the domain shift of SSL encoding phonetic distributions among different age groups. 
Fine-tuned SSL models learn to recognize younger children's phoneme distinctions by adapting the phonetics learned from older children. 
For infant VC, we illustrate how pre-training SSL models on home recordings versus adult speech affects how the weighted features learned during fine-tuning. SSL model, pre-trained on large-scale home recordings, preserves key paralinguistic features that are essential for enhancing infant VC.
Our work sheds light on understanding how SSL models process and perceive children's speech and infant vocalizations. 

\section{Acknowledgement}
This research is based upon work [partially] supported by the National AI Institute for Exceptional Education through NSF and IES award \#2229873.

\bibliographystyle{IEEEtran}
\bibliography{refs}

\end{document}